\documentclass[sn-mathphys,iicol]{sn-jnl}

\usepackage{caption}
\usepackage{subcaption}
\usepackage{amsmath}
\usepackage{hyperref}
\newcommand\blfootnote[1]{
  \begingroup
  \renewcommand\thefootnote{}\footnote{#1}
  \addtocounter{footnote}{-1}
  \endgroup
}

\begin{document}

\title{Many-body interactions between contracting living cells}
\markboth{}{}

\author[1]{\fnm{Roman} \sur{Golkov}}\email{romango2@sce.ac.il}
\author[2,3,4,5]{\fnm{Yair} \sur{Shokef}}\email{shokef@.tau.ac.il}
\affil[1]{\orgdiv{Department of Mechanical Engineering}, \orgname{Shamoon College of Engineering}, \city{Ashdod}, \postcode{77245}, \country{Israel}}
\affil[2]{\orgdiv{School of Mechanical Engineering}, \orgname{Tel Aviv University}, \city{Tel Aviv}, \postcode{69978}, \country{Israel}}
\affil[3]{\orgdiv{Center for Physics and Chemistry of Living Systems}, \orgname{Tel Aviv University}, \city{Tel Aviv}, \postcode{69978}, \country{Israel}}
\affil[4]{\orgdiv{Center for Computational Molecular and Materials Science}, \orgname{Tel Aviv University}, \city{Tel Aviv}, \postcode{69978}, \country{Israel}}
\affil[5]{\orgdiv{International Institute for Sustainability with Knotted Chiral Meta Matter}, \orgname{Hiroshima University}, \city{Higashi-Hiroshima}, \postcode{Hiroshima 739-8526}, \country{Japan}}

\abstract{The organization of live cells into tissues and their subsequent biological function involves inter-cell mechanical interactions, which are mediated by their elastic environment. To model this interaction, we consider cells as spherical active force dipoles surrounded by an unbounded elastic matrix. Even though we assume that this elastic medium responds linearly, each cell's regulation of its mechanical activity leads to nonlinearities in the emergent interactions between cells. We study the many-body nature of these interactions by considering several geometries that include three or more cells. We show that for different regulatory behaviors of the cells' activity, the total elastic energy stored in the medium differs from the superposition of all two-body interactions between pairs of cells within the system. Specifically, we find that the many-body interaction energy between cells that regulate their position is smaller than the sum of interactions between all pairs of cells in the system, while for cells that do not regulate their position, the many-body interaction is larger than the superposition prediction. Thus, such higher-order interactions should be considered when studying the mechanics of multiple cells in proximity.}

\maketitle

\blfootnote{We dedicate this article to Fyl Pincus, who promoted the field of soft matter forward, both by his own scientific achievements, and more importantly by him pushing and encouraging young scientists in the field.}

\section{Introduction}

Live cells exert contractile forces on their environment. The shape, size, and resulting biological function of each cell are determined by the balance of internal and external mechanical forces applied on the cell's surface, such as polymerization or contraction of cytoskeletal networks, changes in internal osmotic pressure, or forces exerted on the cell by its neighbors~\cite{Paluch_2009}. Actomyosin networks within living cells generate and transmit these forces to the extracellular matrix (ECM) via focal adhesions~\cite{Balaban_2001}. The resulting balance of forces is regulated by the cell and may change in response to changes in the rigidity of the ECM~\cite{Eastwood1998, Tee_2010, Schwarz_Safran_RMP2013}. It is not fully clear how cells respond to changes in the mechanical environment caused by other cells, external forces, or changes in the rigidity of the medium. In many studies, the working hypothesis has been that cells tend to maintain specific quantities through mechanical homeostasis~\cite{De_Zemel_2008, BenYaakov_Soft_Matter_2015}. For example, by regulating the forces they apply, cells will vary the displacements they generate as their environment changes. Alternatively, cells may change the forces needed to create those displacements by regulating their deformation. Furthermore, cells modulate their shape and spatial contractility patterns in response to environmental changes. 

The mechanical activity of cells is often described by force dipoles, namely pairs of equal and opposite active forces that each cell applies on its mechanical environment~\cite{Schwarz_PRL_2002, Bischofs_PRE_2004, Bischofs_PRL_2005}. There are analogies between such force dipoles and electric dipoles that consist of two equal and opposite electric charges. Similarly, mechanical interactions between cells result from each cell generating a deformation field in the surrounding medium, which resembles the electric field formed around an electric dipole. Distant cells are, in turn, influenced by this field. Thus, matrix-mediated interactions between cells are similar but not identical to interactions between electric dipoles. 

\begin{figure}[b]
\centering
\includegraphics[width=0.4\columnwidth]{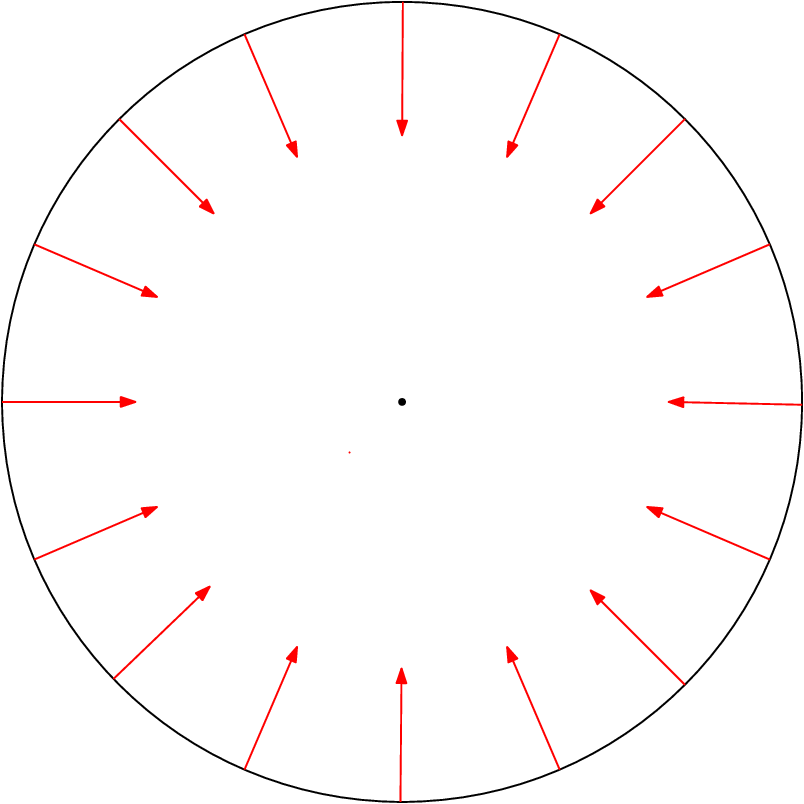}
\caption{Each cell is modeled as a spherical force dipole, comprised of radial active forces that are isotropically distributed on the surface of a sphere}
\label{FD}
\end{figure}

A tractable approach to theoretically describe such contractile cells, which will be employed here, is by modeling them as spherical force dipoles~\cite{Shokef_PRL_2012, BenYaakov_Soft_Matter_2015, Xu_PRE_2015, Sirote_PRE_2021}, i.e., spherical bodies that apply isotropic contractile forces on their surrounding matrix, as depicted in Fig.~\ref{FD}. The mechanical response of the ECM is strongly nonlinear~\cite{Gardel_2004, Storm_2005, Vader_2009}, which has many implications on matrix-mediated elastic interactions between cells~\cite{Winer_PLOS_ONE_2009, Shokef_PRL_2012, Xu_PRE_2015, Ronceray_2016, Sopher_2018, Ronceray_2019, Goren_2020, Sirote_PRE_2021, Mao_Shokef_editorial}. Nonetheless, one can study the elastic interaction between spherical cells surrounded by a linearly elastic material~\cite{BenYaakov_Soft_Matter_2015, Golkov_NJP_2017, Sirote_PRE_2021}. The concepts introduced and the physical mechanisms identified in such studies are also relevant to morphologically complex cells in nonlinear materials. Specifically, despite the linear properties assumed for the ECM, the intra-cellular mechanisms for regulating each cell's mechanical activity give rise to nonlinearities that show up in inter-cellular behavior. 

In this paper, we investigate how cellular regulation of mechanical activity breaks the superposition that one could naively expect to find due to the linear elastic response of the surrounding medium. We analyze situations containing multiple contractile cells and show that the total interacting energy in such cases differs from the result obtained by assuming that the interactions are pairwise additive.

\section{Shape Regulation}

We distinguish between two types of spherical force dipoles, based on the presence or absence of regulation of the forces that they apply; \emph{dead}, but active force dipoles do not regulate the forces that they apply, and their activity does not depend, for instance on the distances to their neighbors. In our model, \emph{live} cells are capable of measuring external forces and deformations on their surface and adjusting the active forces that they apply according to some internal algorithm, for example, to maintain a certain displacement or a certain force on their surface. 

\begin{figure}[bth]
\centering
\begin{subfigure}{0.8\columnwidth}
\includegraphics[width=\columnwidth]{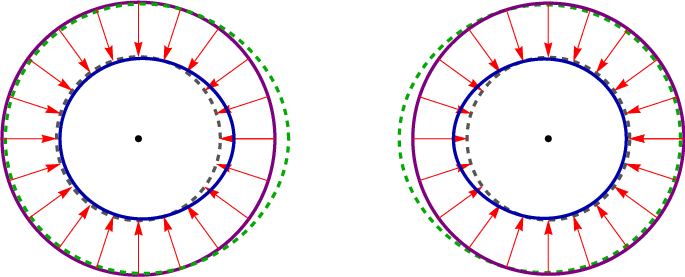}
\caption{Dead}\label{Dead}
\end{subfigure}
\begin{subfigure}{0.8\columnwidth}
\includegraphics[width=\columnwidth]{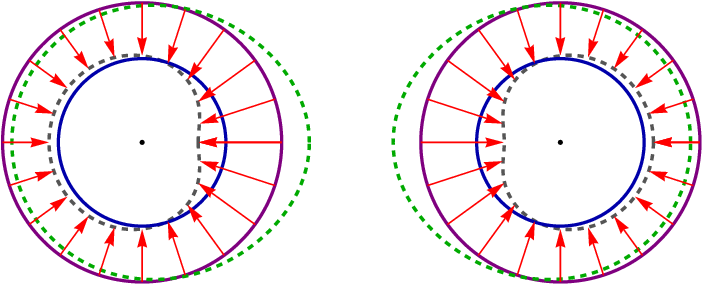}
\caption{Live, fixed size  fixed position}
\label{LiveFP}
\end{subfigure}
\begin{subfigure}{0.8\columnwidth}
\includegraphics[width=\columnwidth]{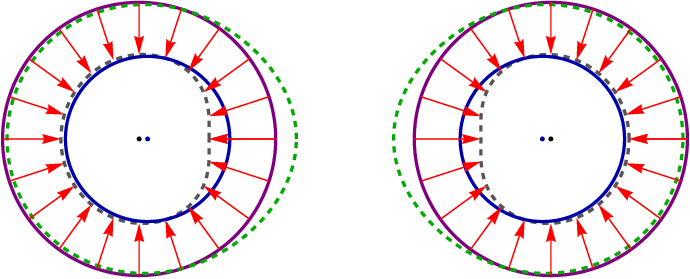}
\caption{Live, fixed size variable position}
\label{LiveVP}
\end{subfigure}
\caption{Two spherical force dipoles: (a) Dead force dipoles applying an isotropic elastic force, (b-c) Live force dipoles regulating the force they apply to remain spherical even in the presence of the other force dipole. Initial shape (purple), forces applied by each force dipole (red arrows), corresponding self displacements (dashed black), displacements caused by the other force dipole (dashed green), total displacement (blue), the center of the interaction-free displacement (black dot), the center of the displacements with the interactions (blue dot). For illustration purposes, the initial distance between the spheres was set to $d = 3 R_0$, and the self-displacement to $u_0 = 0.4R_0$, where $R_0$ is the radius of each sphere. The Poisson ratio is $\nu=0.3$. See text for description of the different regulation scenarios in (b-c)}
\label{FP-VP-two-cells}
\end{figure}

The difference between these two types of behavior is evident when a spherical cell generates a radial and isotropic self-displacement field, i.e., the displacements induced by this cell in the absence of neighboring cells. Such a field would cause the cell to only change its volume, without any distortion of its shape. In that case, 
a pair of such dead active force dipoles preserves their self-displacement fields, leading to vanishing interaction energy~\cite{Golkov_NJP_2017}. Note that if the self-displacement field generated by each sphere is radially symmetric around the center of that sphere, then the total displacement on the surface of each cell, which is the sum of the self-displacement fields generated by these two contracting spheres, would be anisotropic, as shown in Fig~\ref{FP-VP-two-cells}a. This result is general for contracting objects or arbitrary shapes, that generate self-displacements only in their principal directions~\cite{Sines, Mura}.

In comparison, two live cells that adjust the self-displacement fields that they generate have non-vanishing interaction energy. We focus on live cells with shape regulation~\cite{Golkov_NJP_2017}, as demonstrated in Fig.~\ref{FP-VP-two-cells}b,c. Namely, spherical cells that adjust the anisotropic azimuthal distribution of the active forces that they apply, such that the total displacement on their surface will be radially symmetric. This total displacement is the sum of the self-displacement that each cell generates on its surface plus the displacement fields on its surface due to the activity of the other cells in the system. 

\section{Many-body Interactions}

We consider a series of identical live spherical cells of radius $R_0$, arranged along a straight line, separated by equal distances $d$ between their centers, and surrounded by a three-dimensional linear elastic material of bulk modulus $K$ and shear modulus $G$. In the absence of other cells, each cell contracts isotropically with a displacement $u_0$ on its surface. We assume that each cell senses the displacements created on its surface by all other cells and adjusts its active force to compensate for them and prevent distortions of its spherical shape. For this complex situation of multiple activity-regulating cells, we will calculate the total elastic energy stored in the surrounding medium. We will then obtain the many-body interaction energy by subtracting from the energy of this mutual situation, the sum of the self-energies of the cells, namely the energy stored in the medium assuming that each cell contracts independently in an infinite medium, without any other cells. As we will show below, due to the nonlinearity in the regulation of the active forces that the cells apply, this many-body interaction energy differs from the sum of all two-cell interactions. We develop a formalism for an arbitrary number of cells, and will explicitly solve the geometries including three or four cells, and compare them to the pair-wise additive result obtained from the analysis of geometries of two cells. We will also consider an infinite array of equally spaced cells along a straight line, for which we will calculate the interaction energy per cell.

\section{Displacements Created by Spherical Cells}

\begin{figure}[h]
\centering
\includegraphics[width=\columnwidth]{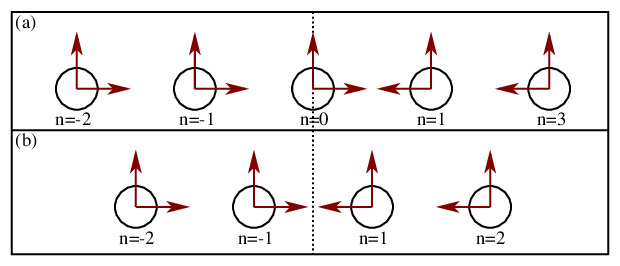}
\caption{Left-handed coordinate systems chosen for cells with positive index $n>0$ and right-handed for zero or negative index $n \le 0$ for (a) odd number of spheres, (b) even number of spheres}
\label{CS}
\end{figure}

\begin{figure}[h]
\centering
\includegraphics[width=\columnwidth]{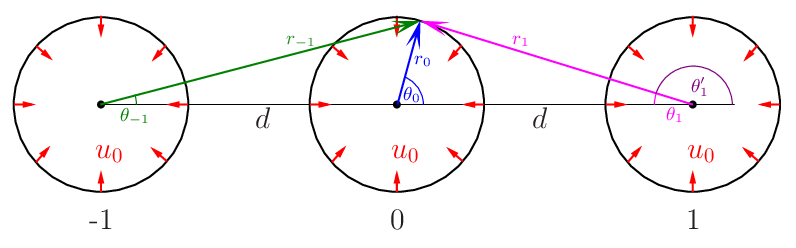}
\caption{Three spherical cells each with radius $R_0$, all applying a radial isotropic displacement $u_0$  (red arrows) on their surfaces. The coordinate systems of spheres $0$ and $-1$ are right-handed (blue and green accordingly) and the coordinate system of sphere 1 is left-handed (magenta) and may be written as $\theta_1=\pi-\theta'_1$ where $\theta'_1$ is the commonly-used right-handed azimuthal coordinate for sphere~$1$}
\label{3cells}
\end{figure}

We describe the displacements generated by cells as the sum of an isotropic constant displacement $u_0$ and an anisotropic, interaction-dependent displacement $\Delta u$ that is intended to cancel anisotropic displacements caused by other cells. The symmetry of the arrangement dictates that the displacement fields produced by cells placed at equal distances on either side of the array's center are mirror images. For simplicity, we place the origin of the coordinate system at this center and number the cells according to their distance from it, see Fig.~\ref{CS}. We choose the coordinate systems of the cells based on their index: left-handed for positive indices and right-handed for negative or zero indices, see Figs.~\ref{CS},~\ref{3cells}. This choice of coordinate systems is based on the system's symmetry and will simplify the calculations. 

The displacement field $\vec{u}$ around each cell must satisfy mechanical equilibrium \cite{Lurie}:
\begin{equation}\label{Navier}
\frac{1}{1-2 \nu} \nabla \nabla \cdot \overrightarrow{u} + \nabla^2 \overrightarrow{u} = 0.
\end{equation}
Due to the rotational symmetry around the line passing through the centers of the cells, there is no dependence on the azimuthal angle $\phi$. Thus, we write Eq. (\ref{Navier}) in spherical coordinates as:
\begin{align}
	&\frac{1}{1-2 \nu} \frac{\partial}{\partial r}  \bigg[ \frac{1}{r^2} \frac{\partial}{\partial r} (r^2 u_r) \nonumber \\ 
	&\qquad\qquad + \frac{1}{r \sin{\theta} } \frac{\partial}{\partial \theta} (u_\theta \sin{\theta})\bigg] \nonumber  \\
	&+ \nabla^2 u_r - \frac{2}{r^2} u_r - \frac{2}{r^2} \frac{\partial{u_{\theta}}}{\partial{\theta}} - \frac{2 u_{\theta} \cot{\theta}}{r^2} = 0, \label{Navier21}\\
	&\frac{1}{1-2 \nu} \frac{1}{r} \frac{\partial}{\partial \theta}  \bigg[ \frac{1}{r^2} \frac{\partial}{\partial r} (r^2 u_r) \nonumber \\
	& \qquad\qquad + \frac{1}{r \sin{\theta} } \frac{\partial}{\partial \theta} (u_\theta \sin{\theta})\bigg]  \nonumber \\
	&\qquad\qquad+\nabla^2 u_{\theta} + \frac{2}{r^2} \frac{\partial{u_{r}}}{\partial{\theta}}-\frac{u_{\theta} }{r^2 \sin^2 \theta}  = 0 , \label{Navier22}
\end{align}
where the Laplacian in spherical coordinates, excluding terms depending on $\phi$, is given by:
\begin{align}\label{Laplacian}
\nabla^2 &= \frac{1}{r^2 \sin \theta} \bigg[ \frac{\partial}{\partial r} \left( r^2 \sin \theta  \frac{\partial}{\partial r} \right) \nonumber \\
&\qquad \qquad \qquad \qquad \quad+ \frac{\partial}{\partial \theta} \left( \sin \theta \frac{\partial}{\partial \theta} \right) \bigg] .
\end{align}

Based on the general solution for the displacement field of a sphere with given cylindrically-symmetric displacements on its surface~\cite{Lurie}, we write the anisotropic displacements field satisfying Eqs. (\ref{Navier21}-\ref{Navier22}) outside the cell ($r>R_0$) as a multipole expansion in terms of spherical harmonics $Y_n(\theta)=\sqrt{\frac{2n+1}{4 \pi}} P_n (\cos{\theta})$:
\begin{align}
&u_{r i} = \frac{u_0 R_0^2}{r_i^2} +  u_0 \sum_{n=0}^{\infty} \left[ n (n+3-4 \nu) \frac{C_n^i R_0^n}{r_i^{n}} \right. \nonumber \\
&\qquad\qquad\qquad\quad\left.-  (n+1) \frac{D_n^i R_0^{n+2}} {r_i^{n+2}} \right] Y_n(\theta_i) , \label{uri} \\
&u_{\theta i} =  u_0 \sum_{n=0}^{\infty} \left[  (-n+4-4 \nu) \frac{C_n^i R_0^n }{r_i^{n}}  \right. \nonumber \\
&\qquad\qquad\qquad\qquad\quad\left. +\frac{D_n^i R_0^{n+2}} {r_i^{n+2}} \right] \frac{dY_n(\theta_i)}{d \theta_i} , \label{uti}
\end{align}
with
\begin{equation}
P_n(x)= 2^n \cdot \sum_{\ell=0}^{n} x^\ell \left(\begin{array}{c} n \\ \ell \\ \end{array}\right)\left(\begin{array}{c} \frac{n+\ell-1}{2} \\ n \\ \end{array} \right)
\end{equation}
the Legendre polynomial of order $n$~\cite{Arfken}. Equations~(\ref{uri}-\ref{uti}) represent the anisotropic displacements created by each cell in its coordinate system with its origin in its center. Here, $u_{r i}$ and $u_{\theta i}$ are the radial and angular components of the displacement field caused by cell $i$, and the infinite sums represent the anisotropic corrections that each cell produces to cancel the shape distortion caused by its neighbors. We have inserted $u_0$ and $R_0$ to make the coefficients $C_n^i$ and $D_n^i$ dimensionless.

Using the dimensionless displacements $\widetilde{u}_{ri}=\frac{u_{ri}}{u_0}$, $\widetilde{u}_{\theta i}=\frac{u_{\theta i}}{u_0}$, and the dimensionless position $\widetilde{r}=\frac{r}{R_0}$, we rewrite Eqs. (\ref{uri}) and (\ref{uti}) as follows:
\begin{align} \label{sphericalharmonicsnormalilzed1}
&\widetilde{u}_{r i} (\tilde{r}_i,\theta_i) = \frac{1}{\widetilde{r}_i^2} + \sum_{n=0}^{\infty} \left[n (n+3-4 \nu) \frac{C_n^i}{\widetilde{r}_i^{n}} \right. \nonumber \\
&\qquad\qquad\qquad\qquad\quad\left.- (n+1) \frac{D_n^i} {\widetilde{r}_i^{n+2}} \right]  Y_n(\theta_i),  \\
&\widetilde{u}_{\theta i} (\tilde{r}_i,\theta_i) =  \sum_{n=0}^{\infty} \left[ (-n+4-4 \nu) \frac{C_n^i}{\widetilde{r}_i^{n}} \right. \nonumber \\
&\qquad\qquad\qquad\qquad\qquad\quad\left. + \frac{D_n^i} {\widetilde{r}_i^{n+2}} \right] \frac{dY_n(\theta_i)}{d \theta} \label{sphericalharmonicsnormalilzed2}
.\end{align}

Note that Eqs.~(\ref{sphericalharmonicsnormalilzed1}-\ref{sphericalharmonicsnormalilzed2}) solve Eq.~(\ref{Navier}) only when each cell is surrounded by an infinite, homogeneous linearly-elastic medium, including in the interior of the neighboring cells. Biological cells have a rigidity that differs from the rigidity of the ECM that surrounds them; thus, this assumption seems problematic. We overcome this by realizing that we may first solve the mechanical problem in which the cells are assumed to have the same linear elastic properties as the ECM. The resultant solution includes a certain stress and displacement on the surface of each cell, and the solution outside the cells is independent of how the cell generates this stress on its surface. In particular, the stress that actual cells apply on their surrounding includes passive stress coming from the rigidity of the cell plus active stress coming from the external forces generated by molecular motors inside the cell. In our analysis, we consider only the total stress and the work it performs, which determines the interaction energy, and our results are valid irrespective of the mechanical rigidity of the cells themselves. See also Ref.~\cite{Sirote_PRE_2021}.

\section{Cancellation Condition}

To preserve isotropic displacements on their surface, live cells in our model create correcting displacements that cancel the anisotropic displacements created by their neighbors. Thus, the sum of all anisotropic displacements caused at the surface of a cell by all other cells and all the corrections applied by the discussed cell must vanish. The coefficients $C_n$ and $D_n$ in Eqs.~(\ref{uri}-\ref{uti}) are derived in this way so that each cell can retain its spherical shape despite interacting with its neighbors. To apply the cancellation condition and to derive from it the expressions for $C_n$ and $D_n$, we transform the expressions for the displacement fields of each cell $j$ to the coordinate system of the discussed cell $i$ by substitution of the expressions for $r_j$ and $\theta_j$ in terms of $r_i$ and $\theta_i$ and then multiplying the displacement vector $\overrightarrow{u_j}=\left(u_{rj},u_{\theta j}\right)$ by a transformation matrix. The transformation matrix depends on the coordinate systems of the cells $i$ and $j$; we use the rotation matrix
\begin{equation}
\textbf{B}_{s}=
\left(
\begin{array}{cc}
\cos \left(\theta _i-\theta _j\right) & \sin \left(\theta _i-\theta _j\right) \\
-\sin \left(\theta _i-\theta _j\right) & \cos \left(\theta _i-\theta _j\right) \\
\end{array}
\right)
\end{equation}
for $i$ and $j$ with the same signs, and the reflection matrix 
\begin{equation}
\textbf{B}_{o}=
\left(
\begin{array}{cc}
-\cos \left(\theta _i+\theta _j\right) & \sin \left(\theta _i+\theta _j\right) \\
\sin \left(\theta _i+\theta _j\right) & \cos \left(\theta _i+\theta _j\right) \\
\end{array}
\right)
\end{equation}
for indices with opposite signs.

The central cell $i=0$ may be treated as having a positive or a negative sign and right or left-handed coordinate system, accordingly. In our analysis, we chose to treat the central cell as having a left-handed coordinate system, and thus, we treat its index as positive (see Fig.~\ref{CS}).

We write the resultant expressions for the radial and angular displacements caused by each cell $j$ on the surface of cell $i$ in terms of the spherical harmonics of cell $i$ by writing the projections:
\begin{align}
\left(u_{r} \right)_n &= 2 \pi \int_0^{\pi} u_r(\theta) Y_n(\theta)sin{\theta}d\theta,  \label{spheremem1}\\
\left(u_{\theta} \right)_n &= \frac{2 \pi}{n(n+1)} \int_0^{\pi} u_{\theta}(\theta) \frac{Y_n(\theta)}{d\theta}sin{\theta}d\theta.
\label{spheremem2}
\end{align}
As may be seen from Eqs.~(\ref{spheremem1},\ref{spheremem2}), every spherical-harmonic mode of cell $j$ contributes to all the modes on the surface of cell $i$. Finally, we sum the contribution from all cells and find the total displacement in each mode.

The anisotropic displacements caused by all cells $j\neq i$ must be canceled on the surface of cell $i$ by the corrections it applies. We write the dimensionless displacement $\widetilde{u}_{ii}$ \textit{created by cell $i$ on its surface} (namely at $\widetilde{r}_i=1$):
\begin{align}
&\widetilde{u}_{rii}(\theta_i) \equiv \widetilde{u}_{r i} (1,\theta_i)= \sum_{n=0}^{\infty} \Bigl[ n (n+3-4 \nu) C_n^i \nonumber\\
&\qquad\qquad\left.- (n+1) (D_n^i - \sqrt{4 \pi} \delta_{n,0})\right]  Y_n(\theta_i) , \label{surfr1} \\
&\widetilde{u}_{\theta ii}(\theta_i) \equiv \widetilde{u}_{\theta i} (1,\theta_i) \nonumber \\
& \quad =   \sum_{n=0}^{\infty}\left[ (-n+4-4 \nu) C_n^i + D_n^i \right] \frac{dY_n(\theta_i)}{d \theta} . \label{surft1}
\end{align}
The term $\delta_{n,0}$ in Eq.~(\ref{surfr1}) is a Kronecker delta, which represents the isotropic radial displacement created by cell $i$ on its surface without the anisotropic cancellation corrections. This constant term does not depend on changes in the cell's environment. The remaining terms are different modes of additional displacement that this cell creates in response to the displacement field induced on its surface by the neighboring cells. The dimensionless displacement $\widetilde{u}_{ji}$ \textit{created by each cell $j$ on the surface of cell $i$} is:
\begin{align}
&\widetilde{u}_{r ji} (\theta_i)= \sum_{n=0}^{\infty} \sum_{m=0}^{\infty} \left[ f_{nm}^{Cr} (\widetilde{d}_{ji}) C_m^j \right. \nonumber \\
&\qquad\quad \left.+  f_{nm}^{Dr} (\widetilde{d}_{ji}) (D_m^j-\sqrt{4 \pi}\delta_{m,0}) \right] Y_n(\theta_i) \label{surfr2} , \\
&\widetilde{u}_{\theta ji}(\theta_i) = \sum_{n=0}^{\infty} \sum_{m=0}^{\infty} \left[ f_{nm}^{C\theta} (\widetilde{d}_{ji}) C_m^j \right. \nonumber \\
&\qquad\qquad\qquad\qquad\left. +  f_{nm}^{D\theta} (\widetilde{d}_{ji}) D_m^j \right] \frac{dY_n(\theta_i)}{d\theta_i}\label{surft2} ,
\end{align}
where the sum over $m$ originates from the fact that the displacement $\widetilde{u}_j$ created by cell j is given by a multipole expansion (\ref{sphericalharmonicsnormalilzed1}-\ref{sphericalharmonicsnormalilzed2}) with the corrective magnitudes $C_m^j$ and $D_m^j$. The sum over $n$ originates from the fact that after the coordinate transformation when these modes are expressed in terms of the spherical harmonics in the coordinate system of cell $i$, each mode from cell $j$ contributes to all the modes of cell $i$. The functions $f_{nm}^{Cr}(\widetilde{d}_{ji})$, $f_{nm}^{Dr}(\widetilde{d}_{ji})$, $f_{nm}^{C\theta}(\widetilde{d}_{ji})$, and $f_{nm}^{D\theta}(\widetilde{d}_{ji})$ depend only on the dimensionless distance $\widetilde{d}_{ji}=\frac{d_{ji}}{R_0}$ between the cells. However, similarly to the transformation matrices, these functions depend on whether the indices $i$ and $j$ have the same or opposite signs. This follows from the choice of the coordinate systems of the cells that were described earlier, see Fig.~\ref{CS}. If the signs are the same, the functions further depend on the sign of the difference $\lvert i \lvert - \lvert j \lvert $ that indicates the side at which cell $j$ is located relative to cell $i$. Thus we make a distinction between $\left( f_{nm}^{Cr} \right)_l$, $ \left(f_{nm}^{Dr}\right)_l$, $\left(f_{nm}^{C\theta}\right)_l$, $\left(f_{nm}^{D\theta}\right)_l$ and $\left( f_{nm}^{Cr} \right)_r$, $ \left(f_{nm}^{Dr}\right)_r$, $\left(f_{nm}^{C\theta}\right)_r$, $\left(f_{nm}^{D\theta}\right)_r$ for $\lvert j \lvert > \lvert i \lvert$ and $\lvert j \lvert < \lvert i \lvert$, accordingly, with $i$ and $j$ of the same signs, and $\left( f_{nm}^{Cr} \right)_o$, $ \left(f_{nm}^{Dr}\right)_o$, $\left(f_{nm}^{C\theta}\right)_o$, $\left(f_{nm}^{D\theta}\right)_o$for $i$ and $j$ with opposite signs. The expressions for all cases are given in Appendix~\ref{appb}.

\begin{figure}[b]
\centering
\includegraphics[width=\columnwidth]{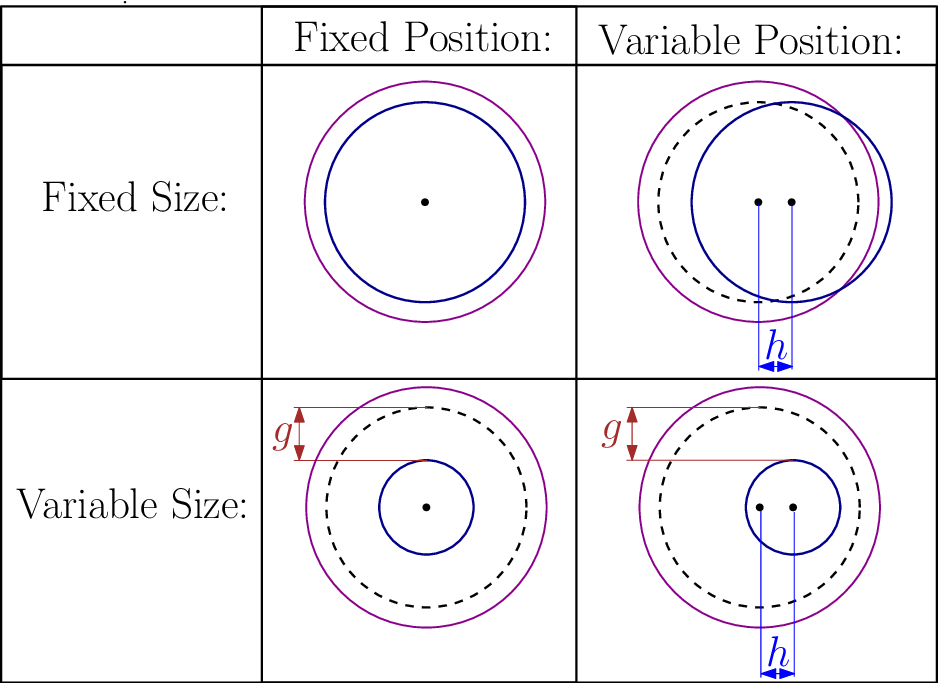}
\caption{Schematic drawings showing how the four possible scenarios of shape regulation depend on whether the size (mode $n=0$) and the position (mode $n=1$) are regulated or not. Initial shape (solid purple), displacement without (dashed black), and with (solid blue) interaction. In the variable size cases, the cell is maintained spherical, yet its radius changes by $g$. In the variable position cases, the center of the cell (dot) translates by $h$}
\label{scenarios_diagram}
\end{figure}

We now require that for live cells, the total displacement $\widetilde{u}_{ii}(\theta_i)+\sum_j \widetilde{u}_{ji}(\theta_i)$ on the surface of cell $i$ is isotropic. We begin by considering the simplest (but strictest) regulation scenario, for which not only is this total displacement isotropic, but its magnitude remains equal to the displacement $u_0$ in the absence of interactions between the cells. Moreover, we require that the center of symmetry of each cell does not move. This will be denoted \emph{fixed size fixed position} (FSFP) regulation. We will also consider three additional activity regulation scenarios in which the interaction causes the cells to change their volume and/or to move, yet they remain spherically symmetric. We denote these regulation scenarios as: \emph{variable size fixed position} (VSFP), \emph{fixed size variable position} (FSVP), and \emph{variable size variable position} (VSVP), see Fig.~\ref{scenarios_diagram} and Fig.~\ref{FP-VP-two-cells} above. For VSFP, cells regulate their shape and rigid body motion, but not their size. In this case, we nullify $D_0$, the first term of the regulating series of each cell responsible for cell size regulation. Similarly, for FSVP, in which cells regulate their shape and size but not their position, we nullify the coefficients $C_1$ and $D_1$, and for VSVP, in which cells regulate their shapes but not their size or position, we nullify $D_0$, $C_1$, and $D_1$. This method is discussed in further detail in Ref.~\cite{Golkov_NJP_2017}.

To preserve isotropic displacement on the surface of cell $i$ we require that:
\begin{align}
\widetilde{u}_{rii}(\theta_i)+\sum_{j\neq i}\widetilde{u}_{rji}(\theta_i)& \equiv 1 , \label{gencancel1}\\
\widetilde{u}_{\theta ii}(\theta_i)+\sum_{j \neq i}\widetilde{u}_{\theta ji}(\theta_i)& \equiv 0 . \label{gencancel2}
\end{align}

Due to the symmetry of the system and our choice of coordinate systems for the cells, the coefficients of pairs of cells with opposite indices are equal, namely $C_m^{j}=C_m^{-j}$ and $D_m^{j}=D_m^{-j}$. Thus, for a system of $k$ cells, we need to write the conditions (\ref{gencancel1},\ref{gencancel2}) for $k/2$ cells with a nonrepeating index $j$ if the total number of the cells is even, and for $k/2+1$ if it is odd.

Substituting (\ref{surfr1},\ref{surft1},\ref{surfr2},\ref{surft2}) in (\ref{gencancel1},\ref{gencancel2}) yields:
\begin{align}
&\sum_{n=0}^{\infty}  \biggl\{\Bigl[n (n+3-4 \nu) C_n^i \nonumber \\ 
&\qquad\qquad- (n+1) (D_n^i - \sqrt{4 \pi} \delta_{n,0})\Bigr] \nonumber\\ 
&\qquad\qquad\qquad\qquad +  \sum_{j \neq i} \sum_{m=0}^{\infty} \Bigl[ f_{nm}^{Cr} (\widetilde{d}_{ji}) C_m \nonumber \\
& + f_{nm}^{Dr} (\widetilde{d}_{ji}) (D_m-\sqrt{4 \pi}\delta_{m,0}) \Bigr] \biggl\} Y_n(\theta_1) = 1 ,\label{fincancel1}\\
&\sum_{n=0}^{\infty}\biggl\{ \left[ (-n+4-4 \nu) C_n^i + D_n^i \right]   \nonumber\\
&\qquad +\sum_{j \neq i} \sum_{m=0}^{\infty} \Bigl[ f_{nm}^{C\theta} (\widetilde{d}_{ji}) C_m + \nonumber \\
&\qquad\qquad\qquad f_{nm}^{D\theta} (\widetilde{d}_{ji}) D_m \Bigr] \biggr\} \frac{dY_n(\theta_1)}{d\theta_1} = 0 . \label{fincancel2}
\end{align}
Due to the orthogonality of the Legendre polynomials, for these infinite sums to satisfy the cancellation conditions, each term in the sums must cancel independently. Thus for all $n \ge 1$ we require:
\begin{align}
	&n (n+3-4 \nu) C_n^i - (n+1) D_n  \nonumber \\
	& \quad +\sum_{j \neq i}\sum_{m=0}^{\infty} \Bigl[ f_{nm}^{Cr}(\widetilde{d}_{ji}) C_n^j  \nonumber \\
	& \qquad\quad +f_{nm}^{Dr}(\widetilde{d}_{ji}) (D_m - \sqrt{4 \pi} \delta_{m,0}) \Bigr] = 0, \label{surf1} \\
	&(-n+4-4 \nu) C_n^i + D_n + \nonumber\\
	& \sum_{j \neq i} \sum_{m=0}^{\infty} \left[ f_{nm}^{C\theta}(\widetilde{d}_{ji})  C_n^j + f_{nm}^{D\theta}(\widetilde{d}_{ji}) D_m \right] = 0. \label{surf2}
\end{align}
Note that for $n=0$, from Eqs.~(\ref{sphericalharmonicsnormalilzed1}-\ref{sphericalharmonicsnormalilzed2}) $C_0$ is irrelevant; thus, we set it to zero. Moreover, since $Y_0(\theta_1)=1$, $\frac{dY_0(\theta_1)}{d\theta_1}=0$ and Eq.~(\ref{fincancel2}) holds trivially, thus for $n=0$ we obtain only one equation, from Eq.~(\ref{fincancel1}):
\begin{align}\label{0eq}
&-D_0 + \sum_{m=0}^{\infty} \Bigl[ f_{0m}^{Cr} C_n^j \nonumber \\
&\qquad\qquad\quad+ f_{0m}^{Dr} (D_m - \sqrt{4 \pi} \delta_{m,0}) \Bigr] = 0 .
\end{align}

We obtain closure of the infinite coupled linear Eqs.~(\ref{surf1}-\ref{0eq}) by assuming that $C_n=0$ and $D_n=0$ for $n>n_{\rm max}$, with some arbitrary value of $n_{\rm max}$, which will determine the accuracy of our calculation. This is justified since we will be interested in large separations between the cells, and since the solutions decay as $1/r^{n}$, at large $r$, large $n$ terms become negligible. We previously verified this numerically by increasing $n_{\rm max}$ until convergence~\cite{Golkov_NJP_2017}. According to our findings, for two cells, $n_{\rm max}=1$ for FP regulation and $n_{\rm max}=2$ for VP regulation scenarios are sufficient to include the leading terms and to obtain good approximations for the interaction energy. Therefore, we use these values of $n_{\rm max}$ also in the present analysis of interactions between multiple cells. The resultant expressions for the coefficients $C_n$ and $D_n$ for three and four cells along a straight line are given in Appendix~\ref{appcoeffs}. We evaluate the forces created by the cell using Eq.~(\ref{taudef}-\ref{sig5}) in Appendix \ref{appc} for the stress tensor in the elastic environment. Due to force balance, the cell's active force per unit area is exactly minus this elastic stress.

The case of many cells along a straight line can be solved by approximating it by an infinite, one-dimensional array of cells; an infinite number of neighbors surrounds each cell. Consequently, all cells respond similarly to their environment and create identical displacement fields. Therefore, $C_n^i=C_n^j$ and $D_n^i=D_n^j$ for any $i$ and $j$. This reduces the number of unknown coefficients from $n_{\rm max}\times k$ in a finite array of $k$ cells to $n_{\rm max}$ in an infinite array, enabling us to define cancellation conditions for a single general cell rather than for $k/2$ cells.

To be finite and solvable, we include in Eqs.~(\ref{gencancel1},\ref{gencancel2}) only terms coming from a limited number $k_{\rm max}$ of neighboring cells, despite the assumption that there is an infinite number of cells. Similarly to $n_{\rm max}$, this is justified since the displacement fields created by the cells decay as $1/r^n$, so displacements produced by distant cells become negligible. As shown below, we verify this numerically by increasing $k_{\rm max}$ until convergence.

\section{Interaction Energy}

We solved the equations for configurations of two, three, four, and an infinite number of cells on a straight line. For each case, we evaluated the extra work performed by each cell $i$ by terminating the infinite sums at $n_{\rm max}=1$ for FP regulation and at $n_{\rm max}=2$ for VP regulation. For configurations that involved three or more cells, we compared the interaction energy obtained from the \emph{direct solution} of the multiple-cell geometry with the pair-wise additive prediction assuming \emph{superposition} of interactions between all pairs of cells within the system. The direct calculation consists of constructing and solving a set of boundary conditions of the form of Eqs.~(\ref{fincancel1}-\ref{fincancel2}). The superposition calculation approximates three-, four-, and many-body interactions by summing all the two-cell interactions in the system. 

The total elastic energy stored in the medium surrounding the cells is equal to the work performed by all cells to generate their deformations, starting from their undeformed states. Cells apply active forces only on their surfaces, thus the amount of work performed by each cell at any point on its surface may be computed by multiplying the force that the cell applies at that point by the total displacement there, divided by two. The division by two results from the integration starting from the undeformed state and reaching the deformed state as the stress in the system gradually builds up linearly with the growing displacement in our linearly elastic medium~\cite{Sirote_PRE_2021}. The self-energy of each cell is the elastic energy it generates when it is surrounded by the infinite ECM and is isolated from other cells. We define the interaction energy as the difference between the elastic energy of the system of cells and the sum of all the cells' self-energies and is thus equal to the extra work performed by the cells due to the presence of other cells around them. 

We write the extra work performed by cell $i$ in a case that includes $k$ cells as:
\begin{equation}
W_i^k = E_0 \widetilde{W}_i^k , 
\end{equation}
where $E_0=8\pi G u_0^2 R_0$ is the cell's self-energy, or the work done by a single, isolated cell that creates on its surface an isotropic displacement $u_0$~\cite{Golkov_NJP_2017}. We find that for all the cases that we considered, the dimensionless extra work may be written as
\begin{equation}\label{dimlessW}
\widetilde{W}_i^k= \frac{ q (1-2 \nu) A_i^k }{ B(\nu) \widetilde{d}^{\alpha} } .
\end{equation}
Here, $A_i^k$ is a numerical prefactor that depends on the number $k$ of cells in the system and on the index $i$ of the cell within the system, but which does not depend on the medium's Poisson ratio $\nu$. We find that this dependence may be included in $B(\nu)$, which is the same for all cells within the system and for any number of cells in the system. Finally, $\alpha$ is the exponent of the power law decay of the interaction energy with the distance between the cells. We find that the sign of the extra work is $q=+1$ for FS and $q=-1$ for VS. These signs are consistent with the theoretical understanding that FS refers to displacement homeostasis, which leads to repulsion between cells and VS to stress homeostasis, which leads to attraction~\cite{BenYaakov_Soft_Matter_2015, Golkov_NJP_2017, Sirote_PRE_2021}. Table~\ref{restab} shows the values of $A_i^k$, $\alpha$, and $B$ for different cells in configurations that include different numbers of cells on a straight line, and for the different position regulation scenarios. Note that by symmetry, $A_j^k = A_{-j}^k$.

For an infinite array of cells, all cells are equivalent. This symmetry cancels the displacements produced by the cell's neighbors, thus its position remains fixed even without position regulation, and position-regulating terms vanish in all regulation scenarios. Thus, the distinction between FP and VP becomes irrelevant. Nonetheless, we refer to the results here as VP regulation, since the positions of the cells are not actively regulated by the cells similar to the VP scenarios in the two, three, and four cell configurations.

\begin{table}[t]
\begin{center}
\caption{Coefficients for additional work Eq.~\eqref{dimlessW} done by the cells as a result of mechanical  interaction with their neighbors, for different scenarios of position regulation.} \label{restab}%
\center
\renewcommand{\arraystretch}{1.5}
\begin{tabular}{| l | c | c | c |}
\hline
  & & FP & VP \\
\hline
  & $\alpha$ & 4 & 6 \\
  & B & $(5-6 \nu)$ & $(4-5 \nu)$ \\
\hline
  Two Cells & $A_{1}^2$ & 1 & 5 \\
  & $A_{\rm tot}^2$ & 2 & 10 \\
\hline
  Three Cells & $A_{1}^3$ & $\frac{5 }{16 } $ & $\frac{685}{64}$ \\
  & $A_0^3$ & $\frac{5}{2} $ & $\frac{45}{4}$ \\
  & $A_{\rm tot}^3$ & $\frac{25}{8} $ & $\frac{1045}{32} $ \\
  & $A_{\rm tot,s}^3$ & $\frac{33}{8}$ & $\frac{645}{32}$ \\
  & $\Delta A_{\rm tot}^3 $ & $-1$ & $\frac{25}{2}$  \\
\hline
  Four Cells & $A_{2}^4$ & $\frac{11 }{162 } $ & $\frac{8510}{729}$ \\
  & $A_{1}^4$ & $\frac{263}{162} $ & $\frac{23635}{1458}$ \\
  & $A_{\rm tot}^4$ & $\frac{274}{81} $ & $\frac{40655}{729} $ \\
  & $A_{\rm tot,s}^4$ & $\frac{2033}{324}$ & $\frac{10165}{324}$ \\
  & $\Delta A_{\rm tot}^4 $ & $-\frac{937}{324}$ & $\frac{71135}{2916}$  \\
\hline
  Infinite Array  & $A_i^\infty$ & & $28.90$ \\
  & $A_{i,{\rm s}}^\infty $ & & $\frac{2 \pi^4}{9}=21.65$ \\
  & $\Delta A_{i}^\infty $ & & $7.35$ \\
\hline
\end{tabular}
\end{center}
\end{table}

We summed these additional works per cell to evaluate the direct interaction energy for a configuration of $k$ cells,
\begin{equation}
E^k = \frac{ q (1-2 \nu) A_{\rm tot}^k E_0 }{ B(\nu) \widetilde{d}^{\alpha} } ,
\end{equation}
where $A_{\rm tot}^k = \sum_i A_i^k$ is also given in Table~\ref{restab}.

\begin{figure}[h]
\centering
\includegraphics[width=\columnwidth]{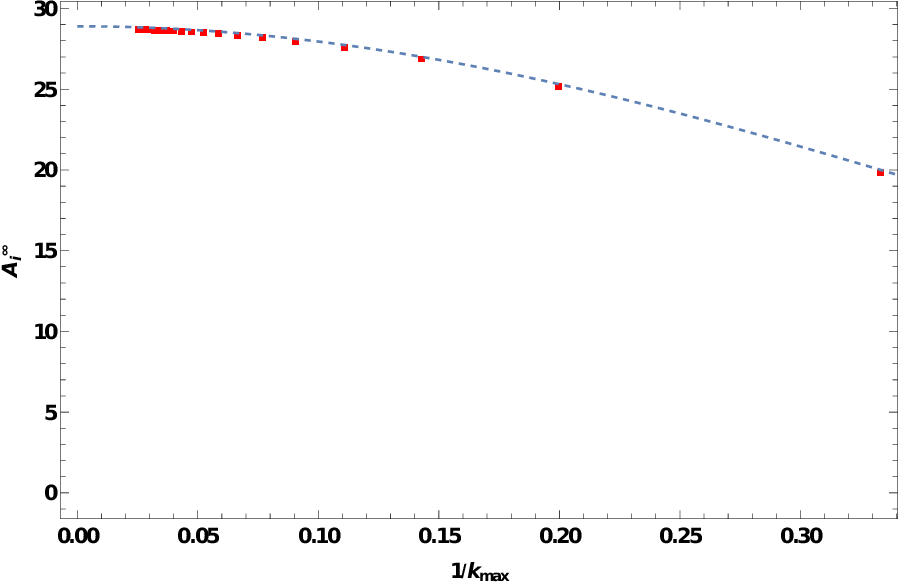}
\caption{Convergence of the coefficient $A_i^\infty$ with the number of cells $k_{\rm max}$ included in the calculation for an infinite array of cells. Exact evaluation (red squares) taking into account interactions of 3 to 39 cells. Approximate fit (dashed line) given by $A_i^\infty=84.42 x^4+28.88 x^3-99.56 x^2+0.1618 x+28.90$ where $x = 1/k_{max}$}.
\label{Ai}
\end{figure}

We find that the sign of the interaction energy, given by $q$, as well as the scaling with distance, given by the exponent $\alpha$, are independent of the number of cells.  Moreover, for the three- and four-cell cases, the additional work performed by the central cells is greater than that performed by the side cells, namely $A_0^3>A_1^3$ and $A_1^4>A_2^4$. This is due to the fact that the central cells are closer to the rest of the cells compared to the side cells. Using the direct method, we did not find a closed-form solution for an infinite array of cells on a straight line. Thus, the number of interacting neighbors included in computations, in this case, is limited. Nevertheless, the addition of interacting neighbors does not influence the sign $q$ of the interaction or the scaling exponent $\alpha$ with cell-cell distances. The energy remains proportional to $\widetilde{d}^{-6}$, and only the coefficient $A_i^\infty$ is affected by the number of neighbors included in the calculation. Figure~\ref{Ai} shows how the value of $A_i^\infty$ converges as the number of interacting cells grows, and the value given in Table~\ref{restab} is for the largest number of cells that we considered, $k_{\rm max}=39$. 

If the interaction energy was pair-wise additive, one could treat the interactions of three, four, and many cells as combinations of two-cell interactions between all the cells in each configuration. The total interaction energy would then be equal to the sum of energies of interactions between all pairs of cells and may be evaluated using the two-cell results given in Table~\ref{restab}. For example, we consider the interaction energy between three cells in the FSFP case by decomposing it into two similar interactions between the side cells and the central cell and the interaction between the cells on opposite sides. Since the distance between these cells equals twice the distance $d$ between the side cell and the central cell, the interaction energy becomes:
\begin{align}\label{supexample}
&\widetilde{W}_{\rm tot,s}^3=\left[ 2 \cdot 2 \frac{1}{\widetilde{d}^4} + 2 \frac{1}{\left( 2 \widetilde{d} \right)^4} \right] \frac{ (1-2 \nu) }{ 5-6\nu} \nonumber \\
&\qquad\qquad\qquad= \frac{33}{8} \frac{ (1-2 \nu) }{ 5-6\nu} \frac{1}{\widetilde{d}^4}
\end{align}

In the same manner, we evaluate the added work performed by a cell in an infinite array of cells as a sum of pair interactions with each cell on both sides, where according to Table~\ref{restab}, in the VP cases, the interaction energy of each pair equals $\frac{10(1-2\nu)}{5-6\nu} \frac{1}{\widetilde{d}^4}$. Thus,
\begin{align}
&\widetilde{W}_{\rm tot,s}^{\infty}=\left[ 2\cdot \sum_{k=1}^{\infty} \frac{1}{\left(k \cdot  \widetilde{d} \right)^4} \right] \frac{10(1-2\nu)}{5-6\nu} \nonumber \\
& \qquad\qquad\qquad =\frac{2 \pi^4}{9}\frac{1-2\nu}{5-6\nu} \frac{1}{\widetilde{d}^4}.
\end{align}

We denote the results obtained from this superposition calculation by the subscript s, and list these results as well in Table~\ref{restab}. We denote the difference between the direct many-body calculation and the superposition expression as $\Delta A \equiv A_{\rm tot}-A_{\rm s}$. 

\begin{figure}[b]
\centering
\begin{subfigure}{\columnwidth}
\includegraphics[width=\columnwidth]{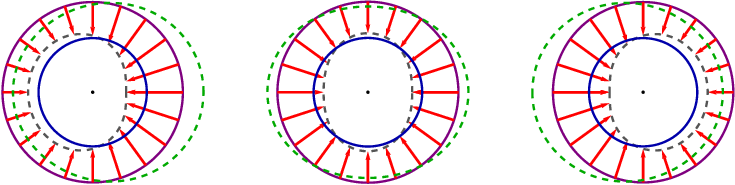}
\caption{Interaction between cells in FP regulation case.}
\label{3FP}
\end{subfigure}
\begin{subfigure}{\columnwidth}
\includegraphics[width=\columnwidth]{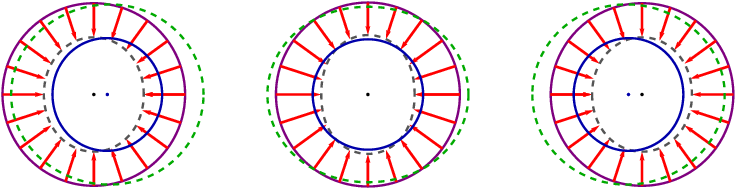}
\caption{Interaction between cells in VP regulation case.}
\label{3VP}
\end{subfigure}
\caption{Interactions between three cells in FP (a) and VP (b) regulation cases. For FP, the cells on the sides apply forces to regulate their motion. Due to the symmetry, the central cell does not need to apply forces to stay in place. For VP, no cell regulates its motion. Initial shape (purple), forces applied by each cell (red arrows), corresponding self displacements (dashed black), displacements caused by the other cells (dashed green), total displacement (blue), the center of the interaction-free displacement (black dot), the center of the displacements with the interactions (blue dot). For illustration purposes, the initial distance between the spheres was set to $d = 3 R_0$, and the self-displacement to $u_0 = 0.4R_0$. The Poisson ratio is $\nu=0.3$}
\label{FP-VP}
\end{figure}

From Table~\ref{restab}, we see that the coefficients $A_i$ found using the two methods for the same configurations are different. This difference is not obvious since, in linear elasticity, typically results may be superimposed when analyzing more complicated arrangements. We conclude that the active response of the cells to their neighbors produces non-linear intercellular interactions, even in the case of linear elasticity. Considering that each active cell in the presence of other active cells is performing extra work, one might expect the interaction energy in all cases to be higher in direct method solutions than in superposition method solutions. However, in three- and four-cell configurations, this assumption is correct in the VP but not in the FP regulation scenarios. Namely, for VP, $A_{\rm tot}>A_s$, while for FP $A_{\rm tot}<A_s$. This unexpected result follows from canceling the central cell's rigid body motion due to the configuration's symmetry. For FP, a large part of the added work comes from the interaction between the size regulation of the cell ($n=0$ mode) and the forces that regulate the motion of the neighbor ($n=1$ mode) as a rigid body. The added work done by the central cell is significant due to its interaction with two neighbors on both sides. The work done by the side cells is small due to the absence of first-mode forces created by the central cell, see Fig.~\ref{3FP}. Most of the added work done by the side cells follows from their interaction with the relatively distant cells on the opposite sides and is small due to the fourth power of normalized distance $\widetilde{d}$. In contrast to FP cases, position-regulating terms are assumed to vanish and are not affected by symmetry in VP cases, see Fig.~\ref{3VP}. The first mode is the only one that includes an antisymmetric function; thus, only this mode will be affected by symmetry.

\section{Conclusions}

We model live cells in the ECM as spherical active force dipoles, which are surrounded by a linear elastic environment. For isotropic active forces and thus isotropic self-displacements, the interaction energy between cells vanishes. Hence, we distinguish between this dead behavior, in which the cells apply constant forces and self-displacements on their surface, and live, regulatory behavior, in which cells adjust their active forces and self-displacements in response to changes that they sense in their environment. This live behavior of cells is similar to interaction between induced electric dipoles on particles with charge regulation.

We examine systems with three, four, and infinite numbers of cells on a line. We solved the interaction energy for these configurations for four different types of self-regulation: on top of preserving their spherical shape, cells can also preserve their volume or their position, or both. Similarly to the interaction between two such shape-regulating cells~\cite{Golkov_NJP_2017}, for fixed position, we found the interaction energy to be inversely proportional to the distance between the cells to the fourth power, and for variable position, to its sixth power. As in the case of two cells, also here, we found that for fixed volume, multiple cells are repelled from each other, and for variable volume they are attracted to each other.  

We compared the results of direct computation of the many-body configurations to the sum of all two-cell interactions for the same configurations. A comparison of the results shows that the superposition method does not predict the energy of multiple-cell configurations. We also found that if cells regulate their position, the many-body interaction energy is smaller than the sum of interactions between all pairs of cells in the system, while for cells that do not regulate their position, the many-body interaction is larger than the superposition prediction. We conclude that the active response of cells to their neighbors produces non-linear intercellular connections even in the case of linear elasticity. 

We have solved the deformation fields for the case in which the rigidity of the cells is the same as that of their environment. Biological cells, however, are complex entities whose rigidity varies from place to place and from the rigidity of the ECM. To relate our results to live cells, we describe each of them as a mechanism that applies forces on the surface and responds by their variation to the application of external force or displacement. The displacements and forces applied by a cell may be divided into ``dead'' and ``live'' parts. While the dead part of the forces or displacements would remain the same if the cells were dead and retained their elastic properties, the live part depends on their programmed behavior and is generated by the contraction of their actomyosin networks. Since the resultant force and displacement are the sum of those two parts, cells may create such a live response so that the resulting forces and displacements will coincide with the case considered here, for which their rigidity is identical to the rigidity of their environment. Even if cells do not behave in this manner, our results highlight the many-body nature of matrix-mediated elastic interactions between cells, and specifically the different behavior for different regulation scenarios.

Following our work on multiple cells along a straight line, it would be interesting to extend our work to two-dimensional arrangements, the simplest of which would be three cells at the corners of a triangle. Since there will be no cylindrical symmetry like in our present study, a more complicated analysis of displacements on the surface of the cells will be needed to model their behavior. It would also be interesting to expand our present work to the case of aspherical cells, for example, oblate spheroids. In this case, the interaction energy between two such cells would depend on the distance between their centers and on the relative angle between their axes. We limited ourselves to cells surrounded by a linearly elastic medium, so that we could exactly solve their interactions analytically. It would be interesting to test our qualitative predictions by solving with numerical simulations situations with nonlinear response of the medium.

\begin{appendices}

\section{The functions $\left(f_{nm}^{Cr}\right)_t$, $\left(f_{nm}^{Dr}\right)_t$, $\left(f_{nm}^{C\theta}\right)_t$ and $\left(f_{nm}^{D\theta}\right)_t$} \label{appb}

We distinct between three different cases for the functions $\left(f_{nm}^{Cr}\right)_t,\ \left(f_{nm}^{Dr}\right)_t,\ \left(f_{nm}^{C\theta}\right)_t$ and $\left(f_{nm}^{D\theta}\right)_t$ appearing in Eqs.~(\ref{surfr2},\ref{surft2}). Thus the index $t$ in the following expressions may be equal to $l$, $r$ or $o$:

\begin{align}\label{f1f2f3f4}
	&\left(f_{nm}^{Cr}(\widetilde{d})\right)_t = {2 \pi}  \int_{0}^{\pi}\bigl[\left(g_m^1\right)_t Y_m\left(\psi\right) \nonumber \\
	&\qquad\qquad\quad +\left(g_m^2\right)_t Y_{m+1}\left(\psi\right)\bigr] Y_n(\theta_1)d\theta_1 , \\
	&\left(f_{nm}^{Dr}(\widetilde{d})\right)_t =  {2 \pi} \int_{0}^{\pi}\bigl[\left(g_m^3\right)_t Y_m\left(\psi\right) \nonumber\\
	&\qquad\qquad\quad+\left(g_m^4\right)_t Y_{m+1}\left(\psi\right)\bigr] Y_n(\theta_1)d\theta_1 , \\
	&\left(f_{nm}^{C\theta}(\widetilde{d})\right)_t = {\frac{\sqrt{\pi(2n+1)}}{n(n+1)}}   \int_{0}^{\pi} \bigl[\left(g_m^5\right)_t Y_m \left(\psi\right) \nonumber\\
	&\qquad+\left(g_m^6\right)_t Y_{m+1}\left(\psi\right)\bigr] \nonumber\\
	& \qquad\qquad\cdot [\cos(\theta_1) P_n(\theta_1) - P_{n+1}(\theta_1)]d\theta_1] , \\
	&\left(f_{nm}^{D\theta}(\widetilde{d})\right)_t =  {\frac{\sqrt{\pi(2n+1)}}{n(n+1)}} \int_{0}^{\pi} \bigl[ \left(g_m^7\right)_t Y_m \left(\psi\right) \nonumber \\
	&\qquad+  \left(g_m^8\right)_t Y_{m+1}\left(\psi\right)\bigr] \nonumber\\
	& \qquad\qquad\cdot [\cos(\theta_1) P_n(\theta_1) - P_{n+1}(\theta_1)]d\theta_1] .
\end{align}
We used the identity \cite{YPn}:
\begin{align}\label{Yder}
	&\frac{dY_n(\theta)}{d\theta}=\sqrt{\frac{2n+1}{4\pi}}(n+1) \nonumber \\
	& \qquad\qquad \cdot \left[\frac{P_{n+1}(\cos \theta )}{\sin \theta }-\cot \theta  P_n(\cos \theta )\right]
\end{align}
to rewrite the derivatives $\frac{dY_n(\theta)}{d\theta}$ in terms of $\theta$, and for the sake of brevity we defined:
\begin{equation}
	\left(g_m^1\right)_t=\left\{ 
	\begin{array}{ll}
		\frac{\left(h_m^1- h_m^2 \right) \sin (\theta_1 )} {{\zeta_t^{{m+1}}}}& for\ t=o,r\\
		\frac{\left(h_m^1+ h_m^2 \right) \sin (\theta_1 )}{ {\zeta_l^{{m+1}}}}& for\ t=l
	\end{array}
	\right.
\end{equation}

\begin{align}
	&h_m^1=\widetilde{d}^2 \left[m^2-m (3-4 \nu )-4 (1-\nu )\right] \nonumber \\
	&\qquad\qquad\qquad\qquad\qquad+m^2 + m (3-4 \nu )
\end{align}
\begin{equation}
	h_m^2=2 \widetilde{d} \left(m^2-2+2 \nu \right) \cos (\theta_1 )
\end{equation}
\begin{equation}
	\left(g_m^3\right)_t =\left\{ 
	\begin{array}{ll}
		\frac{\left( m+1\right) \sin (\theta_1 )}{{\zeta^{m+1}_t}}& for\ t=o,l\\
		-\frac{\left( m+1\right) \sin (\theta_1 )} {{\zeta^{m+1}_t}}& for\ t=r
	\end{array}
	\right. ,
\end{equation}
\begin{equation}
	\left(g_m^4\right)_t =\left\{ 
	\begin{array}{ll}
		\frac{\left( m+1\right) \sin (\theta_1 )} {{\zeta^{m+1}_t}}& for\ t=o,l\\
		-\frac{\left( m+1\right) \sin (\theta_1 )} {{\zeta^{m+1}_t}}& for\ t=r
	\end{array}
	\right. ,
\end{equation}
\begin{equation}
	\left(g_m^5\right)_t=\left\{ 
	\begin{array}{ll}
		\frac{\left(h_m^3-h_m^4\right)}{{\zeta_t^{{m+1}}}}& for\ t=o,r\\
		\frac{\left(h_m^3+h_m^4\right)}{{\zeta_l^{{m+1}}}}& for\ t=l
	\end{array}
	\right.,
\end{equation}
\begin{equation}
	h_m^3=\left(\widetilde{d}^2+1\right) (m+1) (m-4+4 \nu ) \cos (\theta_1 ) 
\end{equation}
\begin{align}
	&h_m^4=\widetilde{d} [ m (m-6+8 \nu )-6 (1-\nu ) \nonumber \\
	&\qquad\qquad\qquad + \left(m^2-2+2 \nu \right) \cos (2 \theta_1 )]
\end{align}
\begin{equation}
	\left(g_m^6\right)_t=\left\{ 
	\begin{array}{ll}
		-(\tilde{d} \cos (\theta_1 )-1) & for\ t=o,l\\
		\cdot \frac{(m+1)(m-4+4 \nu )}{{\zeta^{m}_t}} & \\
		(\tilde{d} \cos (\theta_1 )-1) & for\ t=r \\
		\cdot \frac{(m+1)(m-4+4 \nu )}{{\zeta^{m}_r}} &  
	\end{array}
	\right.
\end{equation}
\begin{equation}
	\left(g_m^7\right)_t=-(m+1) \cos (\theta_1) / {\zeta^{m+1}_t} ,
\end{equation}
\begin{equation}
	\left(g_m^8\right)_t=\left\{ 
	\begin{array}{ll}
		(\widetilde{d} \cos (\theta_1 )-1) & for\ t=o,l \\
		\cdot \frac{ (m+1)}{ \zeta^{m+2}_t} & \\
		- (\widetilde{d} \cos (\theta_1 )-1) & for\ t=r\\
		\cdot \frac{ (m+1)}{ \zeta^{m+2}_r} & 
	\end{array}
	\right.
\end{equation}
where
\begin{align}\label{psizeta}
	\psi_o &\equiv \left[ \widetilde{d}-\cos (\theta_1) \right]/{\zeta_o} , \\
	\psi_l &\equiv \left[ \widetilde{d}+\cos (\theta_1) \right]/{\zeta_l} , \\
	\psi_r &\equiv -\left[ \widetilde{d}-\cos (\theta_1) \right]/{\zeta_r} , \\
	\zeta_o=\zeta_r &\equiv \sqrt{{\widetilde{d}}^2-2{\widetilde{d}}\cos(\theta_1)+1},\\
	\zeta_l &\equiv \sqrt{{\widetilde{d}}^2+2{\widetilde{d}}\cos(\theta_1)+1}.
\end{align}

\section{Coefficients $D_0$, $C_1$, $D_1$, $C_2$ and $D_2$ for large distances}\label{appcoeffs}

Table~\ref{tabcoef3cells} presents expressions for the coefficients $D_0$, $C_1$, $D_1$, $C_2$, and $D_2$ in the case of three cells in a row. Similarly, we present expressions for the same coefficients for four cells in a row in Table~\ref{tabcoef4cells}. In both tables, the coefficients of side cells are marked by index 1 and those of central cells by index 2.
\begin{table}[h]
	\begin{center}
		\begin{tabular}{| c | c | c | c | c | c | c |}
			\hline
			& $FSFP$ & $FSVP$ & $VSFP$ & $VSVP$ \\  \hline
			$n_{\rm max}$ & 1 & 2 & 1 & 2 \\ \hline
			$D_0^1$ & $\frac{5}{16} \frac{\Lambda_1}{\widetilde{d}^4}$ & $-\frac{685    }{64 } \frac{\Lambda_1}{\widetilde{d}^6}$ & 0 & 0 \\[1.3ex] \hline
			$C_1^1$ & $-\frac{15   }{8 } \frac{\Lambda_2}{\widetilde{d}^2}$ & 0 & $-\frac{15   }{8  } \frac{\Lambda_2}{\widetilde{d}^2}$ & 0\\[1.3ex] \hline
			$D_1^1$ & $-\frac{5   }{8  } \frac{\Lambda_2}{\widetilde{d}^2}$ & 0 & $-\frac{5   }{8  } \frac{\Lambda_2}{\widetilde{d}^2}$ & 0\\[1.3ex] \hline
			$C_2^1$ & $-\frac{45   }{32  } \frac{\Lambda_3}{\widetilde{d}^3}$ & $-\frac{45   }{32  } \frac{\Lambda_3}{\widetilde{d}^3}$ & $-\frac{45   }{32  } \frac{\Lambda_3}{\widetilde{d}^3}$ & $-\frac{45   }{32  } \frac{\Lambda_3}{\widetilde{d}^3}$ \\ [1.3ex]\hline
			$D_2^1$ & $-\frac{27   }{16  } \frac{\Lambda_3}{\widetilde{d}^3}$ & $-\frac{27   }{16  } \frac{\Lambda_3}{\widetilde{d}^3}$ & $-\frac{27   }{16  } \frac{3}{\widetilde{d}^3}$ & $-\frac{27   }{16  } \frac{\Lambda_3}{\widetilde{d}^3}$ \\[1.3ex] \hline
			$D_0^0$ & $\frac{5    }{2  } \frac{\Lambda_1}{\widetilde{d}^4}$ & $\frac{45    }{4 } \frac{\Lambda_1}{\widetilde{d}^6}$ & 0 & 0 \\[1.3ex] \hline
			$C_1^0$ & 0 & 0 & 0 & 0\\[1.3ex] \hline
			$D_1^0$ & 0 & 0 & 0 & 0\\[1.3ex] \hline
			$C_2^0$ & $-\frac{5   }{2  } \frac{\Lambda_3}{\widetilde{d}^3}$ & $-\frac{5   }{2 } \frac{\Lambda_3}{\widetilde{d}^3}$ & $-\frac{5   }{2 } \frac{\Lambda_3}{\widetilde{d}^3}$ & $-\frac{5   }{2 } \frac{\Lambda_3}{\widetilde{d}^3}$ \\ [1.3ex]\hline
			$D_2^0$ & $- \frac{3 \Lambda_3}{\widetilde{d}^3}$ & $- \frac{3 \Lambda_3}{\widetilde{d}^3}$  & $- \frac{3 \Lambda_3}{\widetilde{d}^3}$  & $- \frac{3 \Lambda_3}{\widetilde{d}^3}$  \\[1.3ex] \hline
		\end{tabular}
		\caption{Coefficients $C_n^1$, $D_n^1$ (side cells) and $C_n^0$, $D_n^0$ (central cell) of the multipole expansion at asymptotically long distances, $\widetilde{d} \gg 1$ in the case of interaction of three cells. Here $\Lambda_1=\frac{1-2\nu}{5-6\nu}$, $\Lambda_2=\frac{1}{5-6\nu}$, $\Lambda_3=\frac{1}{4-5\nu}$, $n_{\rm max}$ is the highest-order term taken into account, and $\widetilde{d}=\frac{d}{R_0}$ is the dimensionless distance between neighboring cells. } \label{tabcoef3cells}
	\end{center}
\end{table}

\begin{table}[h]
	\begin{center}
		\begin{tabular}{| c | c | c | c | c | c | c |}
			\hline
			& $FSFP$ & $FSVP$ & $VSFP$ & $VSVP$ \\  \hline
			$n_{\rm max}$ & 1 & 2 & 1 & 2 \\ \hline
			$D_0^1$ & $\frac{11}{162 } \frac{\Lambda_1}{\widetilde{d}^4}$ & $\frac{8510   }{729  } \frac{\Lambda_1}{\widetilde{d}^6}$ & 0 & 0 \\[1.3ex] \hline
			$C_1^1$ & $-\frac{49   }{24  } \frac{\Lambda_2}{\widetilde{d}^2}$ & 0 & $-\frac{49   }{24  } \frac{\Lambda_2}{\widetilde{d}^2}$ & 0\\[1.3ex] \hline
			$D_1^1$ & $-\frac{49   }{72 } \frac{\Lambda_2}{\widetilde{d}^2}$ & 0 & $-\frac{49   }{72  } \frac{\Lambda_2}{\widetilde{d}^2}$ & 0\\[1.3ex] \hline
			$C_2^1$ & $-\frac{1255   }{864  } \frac{\Lambda_3}{\widetilde{d}^3}$ & $-\frac{1255   }{864  } \frac{\Lambda_3}{\widetilde{d}^3}$ & $-\frac{1255   }{864 } \frac{\Lambda_3}{\widetilde{d}^3}$ & $-\frac{1255   }{864  } \frac{\Lambda_3}{\widetilde{d}^3}$ \\ [1.3ex]\hline
			$D_2^1$ & $-\frac{251   }{144  } \frac{\Lambda_3}{\widetilde{d}^3}$ & $-\frac{251   }{144  } \frac{\Lambda_3}{\widetilde{d}^3}$ & $-\frac{251   }{144  } \frac{\Lambda_3}{\widetilde{d}^3}$ & $-\frac{251   }{144  } \frac{\Lambda_3}{\widetilde{d}^3}$ \\[1.3ex] \hline
			$D_0^0$ & $\frac{263   }{162  } \frac{\Lambda_1}{\widetilde{d}^4}$ & $\frac{23635    }{1458  } \frac{\Lambda_1}{\widetilde{d}^6}$ & 0 & 0 \\[1.3ex] \hline
			$C_1^0$ & $-\frac{1   }{6  } \frac{\Lambda_2}{\widetilde{d}^2}$ & 0 & $-\frac{1   }{6 } \frac{\Lambda_2}{\widetilde{d}^2}$ & 0\\[1.3ex] \hline
			$D_1^0$ & $-\frac{1   }{18  } \frac{\Lambda_2}{\widetilde{d}^2}$ & 0 & $-\frac{1   }{18  } \frac{\Lambda_2}{\widetilde{d}^2}$ & 0\\[1.3ex] \hline
			$C_2^0$ & $-\frac{275   }{108  } \frac{\Lambda_3}{\widetilde{d}^3}$ & $-\frac{275   }{108  } \frac{\Lambda_3}{\widetilde{d}^3}$ & $-\frac{275   }{108 } \frac{\Lambda_3}{\widetilde{d}^3}$ & $-\frac{275   }{108  } \frac{\Lambda_3}{\widetilde{d}^3}$ \\ [1.3ex]\hline
			$D_2^0$ & $-\frac{55   }{18 } \frac{\Lambda_3}{\widetilde{d}^3}$ & $-\frac{55   }{18 } \frac{\Lambda_3}{\widetilde{d}^3}$ & $-\frac{55   }{18 } \frac{\Lambda_3}{\widetilde{d}^3}$ & $-\frac{55   }{18 } \frac{\Lambda_3}{\widetilde{d}^3}$ \\[1.3ex] \hline
		\end{tabular}
		\caption{Coefficients $C_n^1$, $D_n^1$ (side cells) and $C_n^0$, $D_n^0$ (central cells) of the multipole expansion at asymptotically long distances, $\widetilde{d} \gg 1$ in the case of interaction of four cells. Here $\Lambda_1=\frac{1-2\nu}{5-6\nu}$, $\Lambda_2=\frac{1}{5-6\nu}$, $\Lambda_3=\frac{1}{4-5\nu}$, $n_{\rm max}$ is the highest-order term taken into account, and $\widetilde{d}=\frac{d}{R_0}$ is the dimensionless distance between neighboring cells. } \label{tabcoef4cells}
	\end{center}
\end{table}

\section{The stress tensor}\label{appc}

Here, we develop the expressions for the stress tensor in the cases discussed in the paper. The applied displacements are symmetric about an axis passing through the centers of the cells. The expressions are taken from~\cite{Lurie} for the case of the displacement field given by Eqs.~(\ref{uri}-\ref{uti}), excluding the first term in Eq.~(\ref{uri}), which corresponds to volume change. The stress tensor may be written in the following form in this case:
\begin{equation}\label{taudef}
  \boldsymbol{\tau}=\sum_{n=0}^{\infty} \left(
                                                     \begin{array}{ccc}
                                                       \tau_{rr}^{(n)} & \tau_{r \theta}^{(n)} & \tau_{r \varphi}^{(n)} \\
                                                       \tau_{\theta r}^{(n)} & \tau_{\theta \theta}^{(n)} & \tau_{\theta \varphi}^{(n)} \\
                                                       \tau_{\varphi r}^{(n)} & \tau_{\varphi \theta}^{(n)} & \tau_{\varphi \varphi}^{(n)} \\
                                                     \end{array}
                                                   \right)
\end{equation}
The stress tensor is symmetric ${\tau}_{ij}={\tau}_{ji}$ and thus only six components are to be evaluated. We define the dimensionless stress tensor $\widetilde{\boldsymbol{\tau}}=\frac{\boldsymbol{\tau}}{G} \frac{R_0}{u_0}$, the elements of which are given by:
\begin{align}
	\widetilde{\tau}_{RR}^{(n)} = &2  \bigg[ - \frac{C_n}{\widetilde{r_i}^{n+1}} n (n^2+3n-2 \nu) \nonumber \\
	&+ \frac{D_n}{\widetilde{r_i}^{n+3}}(n+1)(n+2) \bigg] Y_n \left(cos{\theta_i} \right)  \label{sig1} , \\
	\widetilde{\tau}_{R \theta}^{(n)} = &2 \bigg[ \frac{C_n}{\widetilde{r_i}^{n+1}} (n^2-2+2 \nu) \nonumber \\
	&\qquad - \frac{D_n}{\widetilde{r_i}^{n+3}}(n+2) \bigg] \frac{dY_n\left(cos{\theta_i} \right)}{d \theta_i} \label{sig2} , \\
	\widetilde{\tau}_{\theta \theta}^{(n)} = &2 \bigg\{ \bigg[\frac{C_n}{\widetilde{r_i}^{n+1}} n (n^2-2n-1+2 \nu) \nonumber \\
	& \quad - \frac{D_n}{\widetilde{r_i}^{n+3}}(n+1)^2 \bigg] Y_n(cos{\theta_i})   \nonumber\\
	& \quad -  \bigg[ \frac{C_n}{\widetilde{r_i}^{n+1}} (-n+4-4 \nu) + \frac{D_n}{\widetilde{r_i}^{n+3}} \bigg] \nonumber \\
	&\qquad \qquad \qquad  \cdot\frac{dY_n(cos{\theta_i})}{d \theta_i} ctg{\theta} \bigg\} ,  \label{sig3} \\
	\widetilde{\tau}_{\varphi \varphi}^{(n)} =&  2 \bigg\{ \bigg[ \frac{C_n}{\widetilde{r_i}^{n+1}} n (n+3-4n \nu-2 \nu) \nonumber \\
	& \quad-  \frac{D_n}{\widetilde{r_i}^{n+3}}(n+1) \bigg] Y_n(cos{\theta_i})   \nonumber\\
	& \quad+ \bigg[ \frac{C_n}{\widetilde{r_i}^{n+1}} (-n+4-4 \nu) \nonumber \\
	&\qquad\quad+\frac{D_n}{\widetilde{r_i}^{n+3}} \bigg] \frac{dY_n(cos{\theta_i})}{d \theta_i} ctg{\theta} \bigg\} , \label{sig4} \\
	\widetilde{\tau}_{R \varphi}^{(n)} = &\widetilde{\tau}_{\theta \varphi}^{(n)}=0 . \label{sig5}
\end{align}

After obtaining the coefficients $C_n$ and $D_n$, we compute the extra work $W_i^k$  performed by cell $i$ in a configuration that consists of $k$ cells, to generate total displacement $\overrightarrow{u}$ in accordance with Eqs.~(\ref{surfr1}-\ref{surft2}) and the type of the regulation:
\begin{align}\label{inten}
W_i^k &=\frac{1}{2} \int_S \left( \overrightarrow{u} \cdot \overrightarrow{F} - \overrightarrow{u_0} \cdot \overrightarrow{F_0} \right) d s.
\end{align}
Here, the integration is over the spherical surface of the cell $i$, $\overrightarrow{F}$ is the force per unit area applied by it on its environment. Due to force balance, the active forces applied by each cell are equal and opposite to the forces applied on it by the environment: $\overrightarrow{F}= - \boldsymbol{\tau} \cdot \hat{r}$, where $\hat{r}$ is the outward pointing unit vector normal to the surface of the cell after its deformation and movement, and $\boldsymbol{\tau}$ is the stress tensor given above that arises in the elastic environment of each live cell in response to the total displacement $\overrightarrow{u}$ on its surface. Due to the spherical shape regulation of the cells, there are no displacements in the azimuthal direction in this study. $\overrightarrow{F_0}=\frac{4Gu_0}{R_0}\hat{r}$ is the force per unit area on the surface of a single cell with known isotropic displacement $\overrightarrow{u_0} = u_0 \hat{r}$ on its surface and without interactions with other spheres, and thus  $\overrightarrow{u_0} \cdot \overrightarrow{F_0} = 8 \pi G u_0^2 R_0$.
\end{appendices}

\section*{Author Contributions}

RG and YS formulated the problem, performed the calculations, analyzed the results, and wrote the paper.

\section*{Data Availability Statement}

All data is available within the paper.

\bibliographystyle{unsrt}

\end{document}